\documentclass[prb,twocolumn,amsmath,showpacs]{revtex4-1}

\usepackage{graphicx}

\begin{document}

\title{Zero BEC State Amplitude, and BEC Unnecessary to Define Phase $\phi$}
\author{W. M. Saslow}
\email{wsaslow@tamu.edu}
\affiliation{ Department of Physics, Texas A\&M University, College Station, Texas 77843-4242}
\begin{abstract}
We define the ``BEC state'' to be the many-body wavefunction where all particles are in the same one-body state.  Using an argument analogous to Anderson's Orthogonality Catastrophe, we argue that for interacting particles the amplitude of the BEC state within the many-body wavefunction goes to zero in the thermodynamic limit.  This does not mean that there is no condensate.  However, we argue that, if the excitations satisfy the Landau criterion, then the absence of a finite amplitude for the BEC state, or the absence of a condensate, do not prevent the definition of the phase function $\phi$ from which the superfluid velocity follows.  

\end{abstract}

\date{\today}
\maketitle

\section{Introduction}
\label{section:Intro}
It is usually considered that Bose-Einstein condensation (BEC) is necessary for superfluidity in a one-component system.\cite{London,Griffin,LeggettRMP99, LeggettBEC}  BEC is invoked because it enables one to conceptually label a part of the many-body wavefunction (the condensate) that can be associated with a quantum-mechanical phase $\phi$, from which the superfluid velocity is defined via
\begin{equation}
\vec{v}_{s}=\frac{\hbar}{m}\vec{\nabla}\phi,
\end{equation}  
where $m$ is the mass of the particle that constitutes the one-component system.  Certainly BEC is sufficient for $\phi$ to be well-defined, but in the present work we argue that BEC does not occur in the thermodynamic limit, when the number of particles $N\rightarrow\infty$.  However, following Landau's work,\cite{Landau1941} which does not comment on this issue, we argue that BEC is not necessary for $\phi$ to be well-defined.  

In what follows we distinguish between what we call a ``BEC state'' and a ``BEC condensate fraction'' $f_{0}$, and both are distinct from the superfluid fraction $f_{s}$.\cite{Leggett98}  Therefore we consider an issue that, to our knowledge, is new to the literature.  
By BEC state we mean a many-body wavefunction that consists of a product where all of the $N$ particles are in the same one-body state.  (Ref.~\onlinecite{LeggettRMP99} also uses the phrase ``BEC state'', which could also mean the thermodynamic state of a non-interacting Bose system at temperature $T=0$.)  The conceptual advantage of having a finite BEC state is that for this state all of the particles have the same phase (up to a constant).  There can be an infinite number of BEC states, but for a system at rest only the zero-momentum state is relevant. 

On the other hand, the ``BEC condensate fraction'' $f_{0}$ is the {\it fraction} of particles in the one-body state, averaged over the many-body wavefunction.  Thus, with $N_{0}$ the average occupation number for the many-body wavefunction, we have  $f_{0}=N_{0}/N$.  Hence, in the title, by ``average BEC'' we mean ``finite condensate fraction''.  The one-body state may be ``dressed'' by an average one-body interaction with the other particles; it need not be a bare state (such as the zero momentum state).  

This distinction between the BEC state and $f_{0}$ is not merely semantic.  If each configuration included in the many-body wavefunction consists of a product of 80\% particles in the same one-body state, and 20\% particles not in this one-body state, then there is a 0.8 condensate fraction but no BEC because none of the configurations contains 100\% of the particles in the same one-body state.  In this work we argue that the BEC state amplitude goes to zero as $N\rightarrow\infty$.  However, that does not imply that $f_{0}$ goes to zero in that limit.  In fact, the value of $f_{0}$ does not appear in the macroscopic theory of superfluidity.  According to our understanding of Landau's viewpoint, the value of $f_{0}$ is irrelevant to our ability to define $\phi$.  Therefore $f_{0}$ may be zero or it may be non-zero, and we can still (as shown in Sec.~\ref{section:Phase}) define $\phi$.  In fact, for a supersolid, the phase function can depend on many-body correlations, although the simplest form of $\phi$ is the usual one-body form $\phi(\vec{r})$.\cite{SasGalRe}

Our argument that the amplitude of the BEC state, which can be part of a many-body state, has an analog in fermion physics.  Anderson has given a related argument for a non-interacting Fermi system to which a single scattering center is added, sometimes called  the Orthogonalization Catastrophe.\cite{PWA-OC}  He shows that, for $N$ fermions, each non-interacting one-body wavefunction, when perturbed by the scattering center, has nearly unity overlap with the unperturbed scatterer-free states, but the (anti-symmetrized) product of the $N$ wavefunctions has nearly zero overlap with the product of the $N$ scatterer-free states as $N\rightarrow\infty$.\cite{Fermion-OC}  In the present bosonic case, for the unperturbed scatterer-free $N$-body state -- which is precisely the BEC state -- the overlap goes to zero even more strongly (exponentially with $N$) than in the fermionic case (power law in $N^{-1}$).  Hence the amplitude of the BEC state goes to zero as $N\rightarrow\infty$.  

Sec.~\ref{section:Why} gives a detailed argument why the interactions cause the full many-body wavefunction to have essentially zero amplitude for the BEC state, even when there is a finite condensate fraction.   Sec.~\ref{section:Global} discusses how the phase can be defined even when there is no finite amplitude BEC state.  Sec.~\ref{section:Phase} gives a further discussion of the phase.  Sec.~\ref{section:Numerical} proposes a test of the idea that there is a finite superfluid fraction even when the condensate fraction vanishes.  Sec.~\ref{section:Landau} discusses the Landau criterion.  Sec.~\ref{section:Summary} provides a summary and conclusions. 

\section{Why the BEC State has Negligible Amplitude}
\label{section:Why}
First consider the non-interacting Bose system at temperature $T=0$.  The wavefunction in that case is the BEC state.  Although it is Bose-condensed,  with every particle in the same state, as noted by Landau it is unstable to excitations when any object passes through it, and therefore it is not a superfluid.  Bogoliubov found that turning on a weak interaction stabilizes the excitations, so that the system satisfies the Landau criterion for superfluidity.  However, interactions also decrease the condensate fraction, which if BEC were essential to superfluidity, would seem to decrease the stability of the superfluid.  Indeed, in principle it should be possible for the condensate fraction to be zero (not merely zero amplitude of the BEC state), yet if its excitations satisfied the Landau criterion the system would be a superfluid.  Another result of turning on the interaction is that the relative parts of the many-body wavefunction, written in terms of suitably symmetrized products of non-interacting one-body states, have fixed relative phases and fixed amplitude relative to the BEC state, as discussed in detail in Sect. III. 

It might appear that the state found by perturbation theory in Bogoliubov's model has a finite BEC state amplitude.  However, we now argue that is not the case.  Consider an example in which each of the original one-body states within the many-body state, after interactions are included, overlaps by 0.9 with the original common non-interacting one-body state.  (For simplicity it may help to think of the system as having only two available one-body states, with amplitudes 0.9 and $\sqrt{1-(0.9)^{2}}$).  Then for $N$ particles the many-body wavefunction overlaps by $(0.9)^{N}$ with the original non-interacting many-body state, by which we mean the BEC state.  This overlap vanishes in the thermodynamic limit $N\rightarrow\infty$.  Hence there is no many-body BEC, although there is a finite condensate fraction of 0.9.  


This means that, in contrast to the case of a finite amplitude BEC state, there is no obvious label for a state whose phase can be used to define a superfluid velocity.  
The viewpoint of Landau, however, is perfectly compatible with the BEC state having negligible amplitude.  Although Landau does not emphasize the point, the interactions stabilize the wavefunction in the sense that it is a sum of products of one-body terms whose relative amplitude and phase are determined by the interactions via the Schr\"odinger equation.  (We discuss this in detail in Sect. III.)  This makes it possible for the system to have an overall phase function, which is the requirement of Landau's theory of superfluidity.  (Landau's theory is also subject to the stability of the system relative to a moving wall causing spontaneous generation of excitations.)  In principle, many-body effects can modify this phase, but they matter only for systems that are not translationally invariant.\cite{SasGalRe} 


One can argue that it is an oversimplification to begin by describing the interacting system description with unperturbed many-body wavefunctions.  It would be more accurate to consider a Hartree or Hartree-Fock or mean-field approximation to the one-body wavefunctions that includes an average interaction with the other particles.  Those wavefunctions would have a condensate fraction decreased from unity (e.g., 0.7).  If they were {\it exact} eigenstates the system would still be Bose-condensed with a condensate fraction of 0.7.  However, going beyond the average interaction of mean-field theory, the many-body wavefunction becomes more complex on including scattering processes omitted by the average interaction.  If, as in the previous example, scattering causes each of the mean-field one-body states within the many-body state to now overlap, as before, by 0.9 with the common mean-field one-body state, then for $N$ particles the many-body wavefunction now overlaps by $(0.9)^{N}$ with the original common non-interacting many-body state.  The condensate fraction would thus  vanish in the thermodynamic limit.  Hence there is no many-body BEC when interactions are included, although there is a finite condensate fraction of $0.7\times0.9$.  

The above argument does not include higher-order correlations.  However, these should be unimportant for weak interactions, and we expect they would not invalidate the general argument. 

\section{Global Phase is Defined even without BEC}
\label{section:Global}
Consider a system of any number of identical Bose particles. In the absence of interactions, the eigenstates of the system as a whole are products of eigenstates of the individual particles.  The Landau argument applied to the extremely low energy-per-momentum free-particle excitations indicates that the system is not superfluid.\cite{Landau1941}  Now turn on the interactions.  By the fundamental principles of quantum mechanics, any eigenstate of this system can be written as a sum of configurations, each configuration being a product state for the non-interacting system.  By the diagonalization procedure associated with finding eigenstates of the Schr\"odinger equation, each configuration has a well-defined amplitude and phase relative to every other configuration.  Only the overall phase of the system is unknown.  This is true even if the condensate fraction $f_0$ goes to zero.  (It is also true for fermions, and for arbitrary mixtures of particles of different types with differing statistics.)  Thus an overall phase for the system can be defined even without a finite $f_0$.  The argument also applies if there are multiple one-body states with macroscopic occupation numbers. 

To be specific, consider conventional perturbation theory with perturbation $V$, as applied to the ground state $\Psi^{(0)}_{I}$ of an interacting system $I$.  With $\Psi^{(j)}_{nI}\equiv|j,nI\rangle$ denoting the states of the non-interacting system $nI$, and normalized to unity, we have the well-known result that 
\begin{equation}
\Psi^{(0)}_{I}=C\Big(\Psi^{(0)}_{nI}+\sum_{j\ne0}\frac{\langle j,nI|V|0,nI \rangle}{E_{j,nI}-E_{0,nI}}\Psi^{(j)}_{nI}\Big),
\label{PertTh}
\end{equation}
where $E_{j,nI}$ is the energy of the noninteracting state $|j,nI\rangle$ and $C$ is a dimensionless normalization constant.  This result applies both to a many-body system and a one-particle system.  Therefore the relative values of the amplitude and phase between any two noninteracting states, mixed in by the interaction, are completely determined.  

More generally, even beyond perturbation theory, we have
\begin{equation}
\Psi^{(0)}_{I}=C'\Big(A_{0}\Psi^{(0)}_{nI}+\sum_{j\ne0}A_{j}\Psi^{(j)}_{nI}\Big),
\label{PertTh2}
\end{equation}
for some dimensionless $C'$.  Hence, even if, as argued above, there is no finite BEC amplitude (i.e., the dimensionless quantity $A_{0}$ is negligibly small), it is possible to attribute an overall phase to $\Psi^{(0)}_{nI}$, which can be considered to be built into $C$.  If the wavefunction is known analytically, then one can study whether or not such a state is topologically connected.\cite{Leggett1,Kohn}  If the wavefunction is not known analytically, another approach would be to employ Quantum Monte Carlo methods to study the superfluid fraction.\cite{PolCep84,CepRMP}  For a flowing system the phase can be made spatially-dependent, as in the previous section, and one can then study whether or not such a state is stable according to the Landau criterion. 

How to go beyond a one-body phase, to include higher-order correlations, has also been considered.\cite{SasGalRe}  The formalism for the case of a two-body phase has been developed, but not applied, because currently the three-body correlations needed to compute the two-body phase function upper limit are unavailable.  Specifically, the wavefunction with superflow for $i=1,N$ particles can take the form
\begin{eqnarray}
\Psi(\{x_{i}\})&=&\Psi_{0}(\{x_{i}\})\exp(i\Phi), \\
\Phi(\{x_{i}\})&=&\sum_{i}\phi_{1}(x_{i})+\frac{1}{2(N-1)}\sum_{ij}\phi_2(x_{i},x_{j})\cr
&&+\dots
\end{eqnarray}
Here $\Psi_{0}$ is the ground state wavefunction without flow, $\phi_{1}$ is the usual one-body phase function familiar from liquid $^4$He, and $\phi_{2}$ is a new phase function that includes two-body correlations.  The above refers to $T=0$. 

\section{On the Phase of the Field Operator and On ODLRO}
\label{section:Phase}
In many cases the one-body phase $\phi$ associated with superfluidity is considered to be a consequence of the behavior of the two-body correlation function, at a large separation.  Specifically, it is argued\cite{Yang} that one can obtain the phase via the one-body density matrix 
\begin{equation}
\langle\psi^{\dag}(\vec{r}\,')\psi(\vec{r})\rangle\xrightarrow[|\vec{r}\,'-\vec{r}|\rightarrow\infty]{} n_{0}\exp[i\phi(\vec{r}\,')-i\phi(\vec{r})].
\end{equation}
Here $\psi(\vec{r})$ is the one-body field operator, $\psi^{\dag}(\vec{r})$ is its Hermitian conjugate, and $n_{0}$ is the number density of the condensate.  This is known as off-diagonal long-range order,\cite{Yang} or ODLRO.  We believe that this is an assertion rather than a proof that $n_{0}$ can be finite.  Moreover, because the wavefunction is defined only up to an overall phase, the above equation is trivial when associated with a non-zero $n_{0}$.  Because of the Landau argument the phase is defined even when $n_{0}\rightarrow0$.  

In some cases one considers the form $\langle\psi(\vec{r}\,')\rangle$.  Such a field-average is in Fock-space, with many different numbers $M$ of particles.  If the Fock-space states are uncorrelated, then this field-average must be zero.\cite{p39Leggett}  If the Fock-space states are correlated, then the field-average can be non-zero, and is a sum over $M$, with matrix elements having the bra $\langle M+1|$ on the left  and the ket $|M\rangle$ on the right.  As a consequence the matrix element clearly involves integration over the $M$ variables on the right, leaving one variable (call it $\vec{r}_{M+1}$) on the left unspecified, in addition to the variable $\vec{r}$ of the field operator.  

We also refer the reader to Ref.~\onlinecite{LeggettRMP99} for a general discussion of superfluidity, based on the assumption that there is one eigenstate of the one-body density matrix whose eigenvalue is macroscopic.  This assumption is distinct from the assumption that the BEC state has a finite amplitude, but it appears to us that they are likely to be equivalent statements.  Certainly if there were BEC condensation into a single state, one would expect a single eigenstate of the one-body density matrix to have a single macroscopic eigenvalue, and that the corresponding eigenfunction would correspond to the BEC state.  

A similar criticism can be made of ODLRO-based arguments for superfluidity.  Nevertheless, because (following Landau) the overall phase gradient is well-defined, it is unnecessary to use field averages or ODLRO arguments to provide a conceptual basis for the phase.  

Finally, we believe there is no compelling reason why a system cannot have condensate fraction $f_{0}=0$ and yet also have excitations that satisfy the Landau criterion.  In that case one would have a superfluid without a condensate.  We believe that it is worth investigating this possibility in ``analytic'' many-body theories.\cite{Feenberg}


\section{Numerical Test for BEC and Superfluidity}
\label{section:Numerical}
Ref.~\onlinecite{LeggettRMP99} notes that there may be more possibilities for a bose fluid than, as seems to be the case for $^4$He, solidification and BEC condensation.  Bose systems can be simulated numerically, and this suggests a numerical test for the necessity of BEC to superfluidity.  There exist accurate analytic theories for liquid $^4$He that can be used with other potentials, and for which one can expect to obtain reasonably accurate results.\cite{Feenberg}  If one starts with the potential for liquid $^4$He and increases its strength we expect the condensate fraction to decrease.  If one also adjusts the shape of the potential, it may be possible to find potentials that lead to liquid states (at $T=0$) for which the condensate fraction is zero.  There are then two possibilities: either the excitations are stable, or they are unstable (in the sense of the Landau criterion). 

Using such a potential in a quantum Monte Carlo algorithm, the superfluid fraction can be computed via the winding number.\cite{PolCep84,CepRMP}  The necessity-of-BEC viewpoint predicts that if the condensate fraction is zero, then the superfluid fraction must be zero.  The Landau criterion viewpoint predicts that if the excitations are stable then the superfluid fraction must be non-zero.  Such numerical studies should be very revealing.  

\section{On Landau's Critical Velocity Argument: Can Wall Excitations Destroy Superflow?}
\label{section:Landau}
Ref.~\onlinecite{LeggettRMP99} comments that Landau's critical velocity argument most straightforwardly applies only to ions (structureless objects) moving through the putative superfluid.  For example, flow past a wall is more complex than that of an ion because the wall has excitations, and momentum and energy transfer from fluid to wall might involve the wall's excitations.  
In practice, vorticity, rather than ordinary non-topological excitations, is created at the walls, and destroys superfluid flow.  

Nevertheless, consider the possibility that excitations of the wall (rather than the fluid) can take up momentum and energy, and thereby can destroy superfluidity.  The phonons of the wall material, with a finite velocity, will be stable under the Landau criterion.  Since phonons are the only low-energy excitations for insulators, flow past insulators should be stable under the Landau criterion.  

Now consider a confining wall that is a conductor, for which quasiparticles are filled to the top of the Fermi sea.  For flow along $z$ an electron at the Fermi level with momentum along $x$ can absorb momentum along $z$, exciting it to above the Fermi level at the cost of very little energy.  Therefore from energetics alone it appears that interactions of $^4$He with conducting walls can, in principle, destroy superflow.  We are aware of no evidence for this; perhaps the effect is too small to be observed because the electron wavefunctions are exponentially suppressed from ``leaking'' into the $^4$He.\\  

\section{Summary and Conclusion}
\label{section:Summary}
We argue that the BEC state amplitude is zero in the thermodynamic limit of $N\rightarrow\infty$.  We also argue that, if the excitations of the system satisfy the Landau criterion, then a finite one-body condensate fraction is unnecessary for superfluidity, and we propose a test of this proposal by an analytical-numerical many-body theory (to find a liquid state with stable excitations and no condensate) followed by a quantum Monte Carlo calculation to determine the superfluid fraction. 

\section{Acknowledgements}
I would like to thank Alan Griffin for critical conversations and correspondence, Phil Anderson for helpful  discussion, and Matt Sears for critical reading and commentary.  
This work was supported, in part, by the Department of Energy through grant DE-FG02-06ER46278. 

{}


\begin{thebibliography}{}
\bibitem{Fermion-OC} For fermions, because of their higher energy of excitation the states within the Fermi sea are less affected by the perturbation than the states at the Fermi surface.  This leads to a one-body wave-function overlap that is energy-dependent.  The product over all occupied states leads to a many-body overlap varying as a power of $N^{-1}$.
\end{thebibliography}
\end{document}